\documentclass{article}
\usepackage{amsmath}
\usepackage{hyperref}
\usepackage{hyperref}
\usepackage{booktabs} 
\usepackage{caption}    
\usepackage{siunitx}    
\usepackage{multirow}   
\title{Metrics and evaluations for computational and sustainable AI efficiency}
\author{Hongyuan Liu Xinyang Liu Guosheng Hu}
\date{\today}

\begin{document}

\maketitle


\section*{1. Introduction}
The rapid advancement of \textbf{Artificial Intelligence (AI)} has imposed unprecedented demands on \textbf{computational power}.
From model training to \textbf{inference deployment}, each stage of AI computing involves complex mathematical operations and large-scale data processing.
To optimize the \textit{performance}, \textit{efficiency}, and \textit{sustainability} of AI systems, precise measurement and evaluation of their computational processes are critical.

While the community has long emphasized computational metrics such as latency and throughput for service-level objectives (SLOs), a parallel literature has argued for ``Green AI" \cite{schwartz2020green, yigitcanlar2021green, bolon2024review}, elevating energy use and emissions to first-class evaluation criteria alongside task quality. 
The previous researches offer valuable and practical ingredients: hardware counters and profilers from systems research, benchmark suites from machine learning, and carbon-accounting methodologies from sustainable computing \cite{ogundairo2025using, reddi2020mlperf, carbontracker}.
However, there is still no widely adopted, diverse models covered, rigorous, and end-to-end framework that allows researchers and practitioners to measure, compare, and \emph{jointly} optimize these dimensions across heterogeneous hardware devices, software stacks and precision regimes.
To close this gap, we propose a unified, reproducible methodology for AI model inference that integrates computational and environmental metrics under realistic serving conditions and yields a pragmatic, carbon-aware evaluation.

Computational efficiency in deployed AI services is typically characterized by percentiles of request latency (e.g., p50/p95) and by throughput under target loads. Prior systems work shows that tail latencies (p95 or p99) dominate perceived performance at scale and thus must be managed explicitly, not inferred from means or medians \cite{han2015pcs}. However, most LLM benchmarking still relies on single-request microbenchmarks or steady-state batches, obscuring queuing effects and head-of-line blocking that inflate tail latencies under realistic traffic \cite{patke2024queue}. 

In parallel, “Green AI” and subsequent studies have urged the community to report and reduce energy and carbon costs across the model lifecycle, including inference, which often dominates the deployed production systems \cite{husom2025sustainable}. 
Recent work and industry benchmarking consortia have advanced power/energy reporting, yet carbon accounting varies widely across studies \cite{tschand2025mlperf, desislavov2023trends, li2024review}: some exclude data-center overheads (PUE), others assume fixed grid emissions factors, and few align with emerging software carbon-intensity standards. 
The result is a comparability gap that impedes evidence-based choices of precision, hardware, and runtime.


In this report we therefore ground evaluation in a compact but expressive metric suite and in a measurement protocol that travels across devices. Concretely, we treat latency as the time required to complete a single task and throughput as the rate of tasks completed per unit time, both reported as distributions and paired with workload descriptors such as batch-size and sequence-length statistics; we measure power and integrate it over time to obtain energy \(E=\int P\,dt\) expressed in Watt-hours; and we compute location-adjusted carbon \(C=\mathrm{PUE}\times\kappa\times E\), where \(\kappa\) is the contemporaneous grid carbon intensity. These metrics are instantiated under matched accuracy constraints so that efficiency comparisons remain valid across numeric precisions and runtimes. By avoiding single-number summaries and by normalizing for workload mix, we make results portable from data-center accelerators like GH200/A100 to consumer-level/edge platforms such as RTX 4090 and 3090 and to CPU-only baselines.

To make these measurements actionable, we evaluate \emph{multi-precision} quantized models spanning \texttt{FP32}, \texttt{FP16} and \texttt{INT8} and run \emph{identical} model graphs across mainstream AI software stacks (\texttt{PyTorch}, \texttt{TensorRT}, and \texttt{ONNX Runtime}) on heterogeneous hardware. 
This cross-stack, cross-device design isolates the effects of graph compilation, kernel libraries, and runtime scheduling from those of numeric precision and hardware topology. 
For instance, we align calibration and outlier handling so that \texttt{INT8} results on GH200, A100, RTX 4090 and RTX 3090 are directly comparable, and we annotate measurements with interconnect (NVLink versus PCIe), memory type (HBM versus GDDR), and power-management settings to ensure that differences are interpretable rather than incidental.

By systematically categorizing these concepts and implementing a vendor-agnostic instrumentation layer, the report establishes a \emph{rigorous benchmarking} framework for AI systems that is both reproducible and deployment-grounded. The framework produces decision-ready Pareto frontiers linking accuracy, latency, throughput, energy, and carbon, thereby clarifying when lower precision or alternative runtimes are sustainability-positive without unacceptable quality loss and how those trade-offs shift between hardware classes such as GH200/A100 clusters, RTX 4090/3090 workstations, and CPU-only environments. The accompanying code and scripts are publicly available at \url{https://github.com/herkerser/criterion_quantization}, enabling independent verification and facilitating adoption in both industrial and academic settings.





\section*{2. Latency for Computational Efficiency}
Latency is a fundamental metric used to evaluate the responsiveness of AI systems. 
It plays a pivotal role in ensuring that applications provide timely results, directly impacting user experience and the feasibility of real-time AI deployments.
In this section, we first define latency and explore its various components, including compute, network, and ancillary latencies. 
Next, we discuss the importance of measuring latency, highlighting its impact on user experience, system performance, and application feasibility. 
Finally, we present the methodology for calculating latency in AI systems, emphasizing the practical aspects of measurement in real-world scenarios.

\subsection*{2.1 Definition of Latency}
In AI computing, latency refers to the time interval between when an AI system receives input data and when it produces the corresponding output. 
This duration spans several operational phases, each contributing to the total observed latency in the system \cite{latency0}. 
Latency is computed as the average time taken for a set of inferences. Given $N$ inference runs, the latency $L$ is calculated as:
\begin{equation}
L = \frac{1}{N} \sum_{i=1}^{N} t_i,
\end{equation}

where $t_i$ represents the time for the $i$-th inference, excluding data loading and preprocessing. This formula provides a reliable measure of the system’s operational latency during inference, helping ensure accuracy in performance assessments.

The primary components of latency include compute latency, network latency, and ancillary latencies. 
Compute latency represents the time required for executing the model’s computations. Complex models, such as deep neural networks, often exhibit high compute latency due to the large number of parameters involved.
Network latency is crucial in distributed systems, where data must travel between different system components. 
It significantly impacts performance in cloud or multi-node AI systems.
Ancillary latencies include time spent on tasks such as memory-to-processor data transfer, preprocessing, and post-processing, which also contribute to overall system delays.

Latency is often characterized by three primary measures in networked AI environments \cite{latency1}. Head latency represents the minimum observed latency, providing a baseline for optimal network response times. It reflects the best-case scenario under ideal conditions. Average latency is the mean latency observed over a period of time and serves as a general benchmark for system performance. On the other hand, tail latency is concerned with the worst-case latency, often quantified through higher percentiles (e.g., p95 or p99). Tail latency is particularly critical in AI systems because even small delays at the tail can disrupt processing pipelines, affecting overall system efficiency, especially for real-time applications.



\subsection*{2.2. Significance of Latency Measurement}
Latency measurement is critical for several reasons and proves essential for AI systems~\cite{latency0}. First, it directly affects user experience. High latency results in sluggish system responses, harming user satisfaction, especially in real-time applications. Conversely, low latency is essential for providing real-time interaction in systems like conversational AI and autonomous driving.

In mission-critical applications, such as autonomous vehicles or fraud detection, excessive latency can lead to system failure or safety risks, making strict latency limits essential.
Latency measurement is also key for system optimization. Identifying latency bottlenecks enables engineers to focus optimization efforts on the most critical areas, improving efficiency. Additionally, reducing latency enhances cost efficiency by minimizing idle resources and maximizing hardware utilization, particularly in cloud environments where resources are billed.



\section*{3. Throughput in AI Computing}
Throughput is another critical metric for evaluating the computational efficiency and scalability of AI systems. While latency measures how quickly a single request is processed, throughput indicates the system’s ability to handle a high volume of requests over time. In this section, we define throughput and its measurement units, discuss its significance for AI system performance, and analyze how throughput interacts with other metrics such as batch size and latency.

\subsection*{3.1. Definition of Throughput}
Throughput quantifies the rate at which a system processes tasks within a defined timeframe. 
It can be calculated as the ratio of the batch size $B$ to the average latency $L$, as shown in the formula:
\begin{equation}
    \text{Throughput} = \frac{B}{L}
\end{equation}

This relationship highlights how throughput depends on both the number of tasks processed per unit time and the latency of each task. Optimizing throughput requires careful management of batch sizes and latency, ensuring that both metrics are balanced to meet the system’s performance goals.

For AI systems, it is often measured in terms of Requests Per Second (RPS), Transactions Per Second (TPS), or task-specific units like images per second or tokens per second for large language models (LLMs) \cite{throughput0}. In networked environments, throughput can also be expressed in bits per second (bps).

While throughput is often compared to bandwidth, it is essential to understand the distinction between these two concepts. Bandwidth refers to the maximum rate at which data can be transmitted over a communication channel, such as a network connection or memory bus. It is a theoretical upper bound based on the physical capabilities of the medium, typically measured in bits per second (bps) or gigabits per second (Gbps).

On the other hand, throughput measures the actual rate at which data is successfully transmitted or processed by the system, considering factors such as network congestion, system bottlenecks, and hardware limitations. Throughput, therefore, reflects real-world performance, while bandwidth represents the system's capacity in an ideal scenario. In many cases, throughput will be lower than bandwidth due to real-world inefficiencies, such as network latency, packet loss, or processing delays.


    
    

\subsection*{3.2. Significance of Throughput Measurement}



Throughput is a critical metric for evaluating the scalability and efficiency of AI systems, as it indicates the system's ability to handle increasing workloads or concurrent tasks. Proper measurement helps ensure the system remains responsive and stable under peak loads, supporting the performance requirements of large-scale applications.

Throughput also plays a crucial role in resource optimization and cost efficiency. By assessing throughput, systems can better utilize computational resources like GPUs and CPUs, minimizing underutilization and avoiding overload. This leads to reduced operational costs, particularly in cloud environments where resources are metered.

Additionally, throughput directly impacts user experience as well as system health. High throughput enables faster task processing, improving responsiveness and user satisfaction. It also serves as an indicator of system robustness, as systems with higher throughput are better equipped to maintain performance during failures or disruptions, ensuring continued reliability and service quality.

\subsection*{3.4. Latency-Throughput Tradeoff}
There is a fundamental tradeoff between latency and throughput in AI workloads. Larger batch sizes generally increase throughput by processing more data at once, but this can also raise latency due to delays in filling or processing the batch \cite{throughput2}. This tradeoff becomes significant when balancing the needs of real-time applications, which require low latency, against batch processing tasks that benefit from higher throughput.

The batch size has a critical impact on both metrics. 
Static batching is ideal for predictable workloads, such as document processing, maximizing throughput while potentially increasing latency \cite{throughput3}. 
Dynamic batching is more suited for interactive applications, where latency is crucial but throughput needs to be high as well \cite{throughput5}. 
In general, there are diminishing returns when increasing batch size beyond a certain point, especially when hardware limitations, like memory bandwidth, come into play \cite{throughput2}.
For example, GPUs with higher memory bandwidth, such as the H100, may handle larger batches more effectively than older models like the A100.

\section*{4. Environmental Impact: Energy Consumption and Carbon Footprint}

As AI models continue to grow in size and complexity, their environmental impact—particularly in terms of energy consumption and carbon footprint—has become an urgent concern. The computational demands of modern AI systems, especially deep learning models, have significantly increased over the past decade, leading to a corresponding rise in energy consumption and associated carbon emissions. This section provides an overview of the definitions and calculations of energy consumption and carbon footprint, explores the significance of these environmental metrics, and discusses the methods for quantifying and mitigating AI-related environmental impacts.

\subsection*{4.1. Definition of Energy Consumption and Carbon Footprint}
Energy consumption refers to the total amount of electrical energy required for the operation of AI systems. It is typically measured in Watt-hours (Wh), reflecting the amount of energy consumed by the system over time. The energy consumption can be calculated by integrating the power usage over the operational period, as follows:

\begin{equation}
   E=\int P\,dt,
\end{equation}
where $E$ represents the total energy consumed, $P$ is the instantaneous power usage of the system, and $t$ is time. This calculation provides a direct measure of the energy utilized by AI systems during tasks such as model training, inference, and data storage.

Carbon footprint is a measure of the environmental impact in terms of greenhouse gas emissions associated with the energy consumption of AI systems. 
This metric quantifies the amount of CO\textsubscript{2}-equivalent emissions generated by the electricity consumed by the AI system.
The carbon footprint is calculated by considering the Power Usage Effectiveness (PUE), which accounts for the total energy required by the infrastructure (including cooling and networking) relative to the energy consumed by the AI models themselves, and the carbon intensity of the energy used, which varies by region and the energy mix. 
The carbon footprint is calculated as:
\begin{equation}
    C=\mathrm{PUE}\times\kappa\times E.
\end{equation}

In this equation, $C$ represents the carbon footprint in kg CO\textsubscript{2}-equivalent (CO\textsubscript{2}e), where PUE is the Power Usage Effectiveness, and $\kappa$ is the carbon intensity of the electricity used, typically measured in kg CO\textsubscript{2} per kWh. The energy consumption $E$ is obtained from the power usage integral, providing a comprehensive measure of the system’s environmental impact.

These metrics are essential for assessing and comparing the environmental footprint of AI models, particularly as AI applications expand in both size and scope.

\subsection*{4.2. Significance of Environmental Impact Measurement}
Measuring the environmental impact of AI systems, specifically through energy consumption and carbon footprint metrics, is crucial for several reasons. First, it provides a clear understanding of the operational costs associated with AI deployments. As AI models, particularly deep learning models, become larger and more computationally intensive, their energy demands have increased dramatically. For example, between 2012 and 2018, the computational requirements for training deep learning models grew by 300,000-fold, leading to a corresponding surge in energy usage~\cite{carbontracker}.

Second, the carbon footprint of AI systems is closely tied to the energy sources powering data centers. Data centers, which are responsible for hosting and running AI models, consume a significant portion of global electricity with projections indicating that this could rise to 9\% of U.S. electricity consumption by 2030~\cite{future}. The carbon footprint of AI systems is influenced by the grid carbon intensity ($\kappa$), which varies depending on the mix of renewable versus fossil fuel-based energy sources in a given region. By quantifying energy consumption and carbon emissions, it becomes possible to assess and reduce the environmental impact of AI systems, especially by transitioning to renewable energy sources and improving operational efficiency.

Furthermore, environmental impact measurement is essential for aligning AI development with sustainability goals. The growing environmental awareness in AI research and industry calls for integrating green AI practices, such as optimizing AI models for energy efficiency and adopting low-carbon infrastructure. By accurately measuring the energy consumption and carbon footprint of AI systems, organizations can make informed decisions about resource usage, energy sourcing, and hardware choices. This also contributes to transparency, enabling organizations to report their environmental footprint to stakeholders and the public.

In the broader context of climate change mitigation, understanding the environmental impact of AI systems provides an opportunity to minimize AI’s effect to global carbon emissions, while still supporting the development of advanced AI technologies. The carbon emissions associated with training and deploying AI models, such as those used in large-scale natural language processing tasks, have become a key area of research in the quest for sustainable AI practices~\cite{howdo, climate}.

\section*{5. experiments}
This section presents a systematic empirical investigation into the performance and efficiency of neural network models, examining the interplay between numerical precision, computational platforms, and model architecture. The experiments are designed to quantify the impact of these factors by evaluating two distinct model families, ResNet for computer vision and OPT for language processing, across a range of industry standard deployment environments. Key metrics, including throughput, latency, energy consumption, and the associated carbon dioxide emissions, are measured to provide a comprehensive assessment of the trade-offs inherent in modern model optimisation.

The selection of numerical precision for a model's weights and activations is a critical determinant of its computational profile. While the standard 32-bit floating-point (FP32) format ensures high fidelity, its substantial memory and processing overheads have encouraged the adoption of lower-precision alternatives such as 16-bit floating-point (FP16) and 8-bit integer (INT8) formats. These offer significant improvements in throughput and reductions in memory footprint, albeit with a narrower dynamic range that requires careful implementation to preserve model accuracy. The landscape of AI development provides specialised platforms to manage this trade-off. Frameworks like PyTorch serve as a flexible environment for model creation, while dedicated inference engines such as NVIDIA's TensorRT are employed to optimise networks through techniques including layer fusion, kernel auto-tuning, and precision quantisation. To bridge these environments, the Open Neural Network Exchange (ONNX) standard facilitates model interoperability, enabling a consistent evaluation across different frameworks and hardware backends.

\begin{table}[htbp]
    \centering
    \captionsetup{justification=centering, width=\linewidth}
    \caption{Performance and Efficiency Metrics for ResNet Models on NVIDIA RTX 3090. All tests were run with a batch size of 100 and an input shape of (3, 224, 224). The CE metric represents the Carbon Dioxide Emission.}
    \label{tab:vision_models}
    \setlength{\tabcolsep}{2pt}
    \begin{tabular}{ll S[table-format=4.2] S[table-format=2.2] S[table-format=1.3] S[table-format=2.2]}
        \toprule
        & & \multicolumn{2}{c}{\textbf{Performance}} & \multicolumn{2}{c}{\textbf{Efficiency}} \\
        \cmidrule(lr){3-4} \cmidrule(lr){5-6}
        \textbf{Model} & \textbf{Platform \& Precision} & {\textbf{Throughput}} & {\textbf{Latency}} & {\textbf{Energy}} & {\textbf{CE}} \\
        & & {(samples/s)} & {(ms)} & {(Wh)} & {(mg)} \\
        \midrule
        \multirow{4}{*}{\textbf{ResNet-18}} & PyTorch, FP16 & 7922.41 & 12.61 & 0.154 & 8.99 \\
        & ONNX, FP16 & 4471.58 & 22.36 & 0.270 & 15.92 \\
        & TensorRT, FP16 & 2492.40 & 40.12 & 0.399 & 23.25 \\
        & TensorRT, INT8 & 3364.51 & 29.72 & 0.206 & 12.01 \\
        \midrule
        \multirow{4}{*}{\textbf{ResNet-50}} & PyTorch, FP16 & 1518.69 & 65.85 & 0.782 & 45.58 \\
        & ONNX, FP16 & 1910.61 & 52.34 & 0.653 & 38.01 \\
        & TensorRT, FP16 & 1703.77 & 58.69 & 0.647 & 37.68 \\
        & TensorRT, INT8 & 3004.10 & 33.29 & 0.297 & 17.29 \\
        \bottomrule
    \end{tabular}
\end{table}

\begin{table}[htbp]
    \centering
    \captionsetup{justification=centering, width=\linewidth}
    \caption{Performance and Efficiency Metrics for OPT Language Models on NVIDIA RTX 3090. The CE metric represents the Carbon Dioxide Emission.}
    \label{tab:language_models}
    \setlength{\tabcolsep}{2pt}
    \begin{tabular}{ll S[table-format=3.2] S[table-format=4.2] S[table-format=2.2] S[table-format=3.2]}
        \toprule
        & & \multicolumn{2}{c}{\textbf{Performance}} & \multicolumn{2}{c}{\textbf{Efficiency}} \\
        \cmidrule(lr){3-4} \cmidrule(lr){5-6}
        \textbf{Model} & \textbf{\& Precision} & {\textbf{Throughput}} & {\textbf{Latency}} & {\textbf{Energy}} & {\textbf{CE}} \\
        & & {(tokens/s)} & {(ms)} & {(Wh)} & {(mg)} \\
        \midrule
        \multirow{2}{*}{\textbf{OPT-125M}} & FP16 & 394.80 & 374.94 & 3.70 & 215.46 \\
        & INT8 (SmoothQuant) & 429.10 & 155.13 & 1.54 & 89.60 \\
        \midrule
        \multirow{2}{*}{\textbf{OPT-1.3B}} & FP16 & 124.26 & 1207.11 & 16.56 & 964.30 \\
        & INT8 (SmoothQuant) & 294.51 & 44.14 & 0.59 & 34.77 \\
        \bottomrule
    \end{tabular}
\end{table}

A detailed analysis of the experimental results, presented in Table~\ref{tab:vision_models} and Table~\ref{tab:language_models}, reveals strong correlations between metrics alongside significant architectural differences in optimisation efficacy. Across all experiments, a predictable inverse relationship exists between latency and throughput, where a reduction in batch processing time corresponds directly to an increase in the number of samples processed per second. Concurrently, both energy consumption and the resultant carbon dioxide emissions exhibit a direct positive correlation with latency, as longer processing times demand greater power consumption. However, the impact of specific optimisation strategies diverges notably between the vision and language domains. For the ResNet models, the choice of runtime environment at FP16 precision yields inconsistent outcomes; the ONNX runtime improves performance for ResNet-50 over the PyTorch baseline, yet degrades it for ResNet-18. The most consistent and powerful enhancement for these vision models comes from INT8 quantisation with TensorRT, which delivers substantial gains across all metrics compared to the FP16 counterparts.

This effect is far more pronounced for the OPT language models, where INT8 quantisation via the SmoothQuant method offers a transformative, rather than incremental, improvement. The reduction in latency and energy consumption for the OPT-1.3B model, for instance, is over 95 percent, a profoundly greater relative improvement than that observed in the vision models. This stark difference likely stems from the distinct computational bottlenecks characterising these architectures. Large language models are frequently memory-bandwidth bound, a constraint that is drastically alleviated by the reduction in model weight size from quantisation. While vision models also benefit, their performance appears to be more constrained by computational limits in certain layers, rendering the gains from quantisation significant yet less revolutionary. In conclusion, while runtime optimisations are beneficial, precision reduction remains the most impactful technique for enhancing efficiency, and its effectiveness is highly dependent on the target model's architectural properties.





\bibliographystyle{plain}
\bibliography{references}
\end{document}